\documentstyle[amsmath,amssymb,12pt,epsf]{article}
\begin{document}
\begin{flushright} 
{CU-TP-1082}
\end{flushright}
\vskip50pt
\begin{center}
\begin{title}
\title{\large\bf Nuclear A-dependence near the Saturation Boundary\footnote{This research is supported in part by the US Department of
Energy.}}\\

\vskip 10pt
{A.H. Mueller\\

Physics Department, Columbia University\\
New York, N.Y. 10027 USA}
\end{title}
\end{center}
\vskip 10pt
\noindent
{\bf Abstract}

The A-dependence of the saturation momentum and the scaling behavior of  the scattering of a small dipole on a nuclear target are studied in the McLerran-Venugopalan model, in fixed coupling BFKL dynamics and in running coupling BFKL dynamics.  In each case,  we find scaling not too far from the saturation boundary, although for fixed coupling evolution the scaling function for large A is not the same as  for an elementary dipole.  We find that $Q_s^2$ is proportional to $A^{1/3}$ in the McLerran-Venugopalan model and in fixed coupling evolution, however, we find an almost total lack of A-dependence in $Q_s^2$ in the case of running coupling evolution.

\section{Introduction}

The study of high density QCD, states or systems where gluon occupation numbers are large compared to one, has become one of the central topics in QCD.  There is a considerable literature whose focus is explaining much of small x and moderate $Q^2$ HERA data[1-4] as well as the general features of hadron production in ion-ion collisions at RHIC[5-11] using high density wavefunctions as the basic ingredient.  In any such description the central parameter is the saturation momentum, $Q_s,$ the scale at and below which occupation numbers are as large as $1/\alpha$ in the light-cone wavefunction.

The energy dependence of the saturation momentum has been widely studied[12-17] and it now appears that one has pretty good control over that dependence in the context of resummed next-to-leading order BFKL dynamics.  Recently there has also appeared some discussion of the A-dependence of the saturation momentum for large nuclei[11-18] beyond the McLerran-Venugopalan model[19] model.  The conclusion has been pretty much the same as for the McLerran-Venugopalan model, that is $Q_s^2(A) \sim A^{1/3}.$

Our purpose here is to revisit the question of the A-dependence of the saturation momentum implementing the BFKL dynamics more fully than has been done so far.  In case fixed coupling BFKL dynamics is used   we confirm that the primary A-dependence of $Q_s^2(A)$ is proportional to $A^{1/3}.$  The result is given in (26). We also confirm that the scattering of a small dipole, of size $1/Q,$ on the nucleus gives a  scattering amplitude which is only a function of $Q^2/Q_s^2(A)$ so long as one is not too far from the saturation boundary. We note, however, that the scaling function is not the same as for the scattering of a  dipole on another elementary dipole. (In the fixed coupling problem it does not appear possible to sensibly talk of dipole-proton scattering in a perturbative context.) The results for the scattering amplitude are given in (27) and (29).

In case a running coupling BFKL dynamics is used the situation changes radically[16,17].  Now the saturation momentum becomes almost completely independent of A at very large rapidities.  Eq.(44) gives our result here for the weak A-dependence which is present.

In both the fixed coupling and running coupling situations we have found it useful to view the BFKL-QCD evolution in an unusual way.  The traditional way to view the scattering of a small dipole on a larger dipole, or on a hadron or nucleus, is to view evolution as part of the wavefunction of the larger object which is then probed by the smaller dipole.  In the present problem it seems more convenient to view the BFKL evolution as part of the wavefunction of the smaller dipole, which  evolution produces gluons (dipoles) of larger scale which eventually interact with the hadron or larger scale.

Finally, because $Q_s^2$ is not very large we do not feel confident in deciding which picture, McLerran-Venugopalan, fixed coupling BFKL dynamics or running couplng BFKL dynamics is more appropriate for RHIC energies and for the LHC heavy ion regime.  It might well be that the McLerran-Venugopalan model  along with a modest amount of fixed coupling evolution is dominant at RHIC energies so that $Q_s^2 \sim  A^{1/3}$ is the appropriate behavior of the saturation momentum.  We would guess that the running coupling regime is likely dominant for protons and small nuclei at HERA energies, and is perhaps important for large nuclei at LHC energies, but this is by no means certain.  It would be very useful to have some numerical calculations which are reliable near the saturation region to try and see what the A-dependence of $Q_s$ actually is in the energy regions covered by the different accelerators.

\section{The semiclassical region (McLerran-\\
Venugopalan model)}

We begin our discussion of the A-dependence near the saturation boundary with a brief review of the McLerran-Venugopalan model[19].  Let $T_N$ be the  amplitude for the scattering of a quark-antiquark dipole of size $\Delta x_\perp \equiv 1/Q$ on a nucleon. Using a normalization where the cross-section for scattering is

\begin{equation}
\sigma_N = 2 T_N
\end{equation}

\noindent one has

\begin{equation}
T_N = {\pi^2\alpha\over 2N_c} x_\perp^2 x G_N
\end{equation}

\noindent where $xG_N$ is the gluon distribution of the nucleon evaluated at scale $Q^2.$  Let $T_A(b)$ be the corresponding amplitude for the scattering of a dipole of size $\Delta x_\perp$ on a nucleus at impact parameter $b.$  Then, the dipole-nucleus cross-section is

\begin{equation}
\sigma_A = 2 \int d^2b T_A(b)
\end{equation}

\noindent with[19,20]

\begin{equation}
T_A=1-exp[-{\bar{Q}_s^2(A)\over 4Q^2}]
\end{equation}

\noindent where

\begin{equation}
\bar{Q}_s^2(A) = 2\sigma_N\cdot 2{\sqrt{R^2-b^2}}\ \rho Q^2={C_F\over N_c} Q_s^2(A)
\end{equation}

\noindent with, as usual[21]

\begin{equation}
Q_s^2(A) = {4\pi^2\alpha N_c\over N_c^2-1}\ 2{\sqrt{R^2-b^2}}\rho x G.
\end{equation}

\noindent $\rho$ is the nuclear density, $\bar{Q}_s$ is the quark saturation momentum and $Q_s$ the gluon saturation momentum.  It should be emphasized that $Q\equiv {1\over \Delta x_\perp}$ defines $Q.\ \  Q$ is not a true momentum variable connected to $x_\perp -$dependence by a Fourier transform.

Eq.(4) has all the basic properties which we wish to note in this section. First of all (4) exhibits geometric scaling[3].  That is $T_A$ is only a function of $Q_s^2(A)/Q^2.  T_A$ exhibits saturation so that $T_A \simeq 1$ when $\bar{Q}_s^2/(4Q^2)\gg 1.$  Finally, when $4Q^2/\bar{Q}_s^2 \gg 1,T_A$ is additive in the various nucleons. That is

\begin{equation}
T_A\ _{_\simeq\atop_{Q^2/Q_s^2\gg 1}}\ \  {\sigma_N\over 2}\ 2{\sqrt{R^2-b^2}}\ \rho,
\end{equation}

\noindent so that there are no non-trivial nuclear effects present when $Q^2/Q_s^2\gg 1.$

\section{Fixed coupling BFKL dynamics}

Now we shall consider the scattering of a dipole of size $\Delta x_\perp = 1/Q$ on a dipole of size $\Delta x_\perp = 1/\mu$ ({\rm where}\ \  $Q^2/\mu^2 \gg 1)$ and on a ``nucleus''  made up of\   A\  such dipoles distributed uniformly in a sphere of radius\   R\   and having density $\rho = A/({4\pi\over 3}R^3).$  Of course this is not a realistic nucleus, since it is not made up of realistic nucleons,  but it is not possible to deal with a  realistic nucleon in the fixed coupling regime.  Since our object is to study A-dependence we must build our nucleus out of individual objects which do make sense for fixed coupling, and there are small dipoles.

In the vicinity of the saturation boundary the amplitude for scattering of a dipole $Q$ on a dipole $\mu$ is given by[16]

\begin{equation}
T_\mu(\mu, b, Q) = T_0(b\mu)\left({Q_s^2\over Q^2}\right)^{1-\lambda_0}\ell n Q^2/Q_s^2
\end{equation}

\noindent where \ b\ is impact parameter and

\begin{equation}
Q_s^2(b,\mu,Y) = a(b\mu) \mu^2[\ell n{1\over \alpha}\alpha^2]^{{1\over 1-\lambda_0}}\ {exp[{2\alpha N_c\over \pi}{\chi(\lambda_0)\over 1-\lambda_0}Y]\over [\alpha Y]^{{3\over 2(1-\lambda_0)}}}.
\end{equation}

\noindent $T_0$ and \ $a$\ are of order one when $b\mu$ is not large. Eq.(8) is valid in the region

\begin{equation}
1\ll \ell n Q^2/Q_s^2 \ll {\sqrt{{4\alpha N_c\over \pi}\chi^{\prime\prime}(\lambda_0)Y}}
\end{equation}

\begin{equation}
\alpha Y\gg{4\pi\over (1-\lambda_0)^2N_c\chi^{\prime\prime}(\lambda_0)} \ell n^21/\alpha
\end{equation}

\begin{equation}
\alpha Y \gg{9\pi\over 16(1-\lambda_0)^3 N_c\chi^{\prime\prime}(\lambda_0)} \ell n^2(\alpha Y)
\end{equation}

\noindent where $\lambda_0$ is defined by $\chi^\prime(\lambda_0) = - {\chi(\lambda_0)\over 1-\lambda_0},$ and $\chi(\lambda) = \psi(1) - {1\over 2} \psi(\lambda)-{1\over 2} \psi(1-\lambda)$ with $\psi$ the logarithmic derivative of the $\Gamma$-function.  For simplicity of notation we now use $Q_s$ to represent the quark saturation momentum.  Since the issue in this and the next section is the functional dependence of $Q_s$ the distinction between $Q_s$ and $\bar{Q}_s$ is not important.

The conventional way of looking at (8) is to view the dipole $Q$ as measuring the gluon distribution of dipole $\mu$ so that

\begin{equation}
{dxG_\mu(x,b,Q^2)\over d^2b} \sim {1\over \alpha} Q^2\left({Q_s^2\over Q^2}\right)^{1-\lambda_0} \ell n Q^2/Q_s^2
\end{equation}

\noindent when $Q^2/Q_s^2 > 1.$  (We note that when $Q^2/Q_s^2$ is on the order of one, ${dxG_\mu\over d^2b} \sim Q_s^2/\alpha.)$  However, one may view the QCD-BFKL evolution in the opposite sense so that one considers the dipole Q to evolve to lower momentum scales,and then (8) gives the probability of finding a gluon (or dipole) at scale $\mu$ in the parent dipole $Q.$  More precisely

\begin{equation}
{d\over dY} \tilde{N}_g(Q,b,Y,\mu) \sim {1\over \alpha}\left({\mu^2\over \tilde{Q}_s^2}\right)^{1-\lambda_0}\ell n(\tilde{Q}^2/\mu^2)
\end{equation}

\noindent is the number of gluons per unit rapidity at scale $\mu$ in the wavefunction of a dipole of size $\Delta x_\perp = 1/Q.$  In the large $N_c$ limit ${dN_g\over dY}$ would also represent the number of dipoles of size $\Delta x_\perp \ge 1/\mu$ in a parent  dipole of size $1/Q.$  In (14) we have introduced $\tilde{Q}_s$ which is defined by

\begin{equation}
{Q_s^2\over Q^2} = {\mu^2\over \tilde{Q}_s^2}.
\end{equation} 

\noindent That is if one writes

\begin{equation}
Q_s^2(b,\mu,Y) = \mu^2 f(b,Y)
\end{equation}

\noindent then

\begin{equation}
\tilde{Q}_s^2(b,Q,Y) = Q^2/f(b,Y),
\end{equation}

\noindent and

\begin{equation}
T_\mu \sim \alpha {d\tilde{N}_g(Q,b,Y,\mu)\over dY}.
\end{equation}

\noindent In Fig.1, the two directions of evolution are pictured.  In the left-hand part of the picture the shaded saturation region for a dipole of size $\mu$ and rapidity \ Y\ is exhibited.  Only the part of the saturation region where $\ell n k_\perp^2/\mu^2 > 0$ is shown.  In the right-hand part of the picture the part of the saturation region having $\ell n Q^2/k_\perp^2 < 0$ is shown for a dipole of size  $Q$ and rapidity \ $Y.$\  For the $\mu$-dipole, evolution proceeds from $y=0$ up to $y=Y,$ while for the $Q-$dipole evolution goes from $\bar{y}\equiv Y-y=0$ to $\bar{y}=Y.$

Now consider the scattering of a dipole $Q$ on a nucleus made up of dipoles of size $\mu$ and\ \underline{nuclear} density $\rho.$  We view the process in the rest frame of the nucleus and in two steps.  In the first step one takes the number density, ${dN_g\over dy}$, of dipoles of size\   $K$\ and having rapidity\   $y$\  in the wavefunction of dipole $Q,$ and in the second step one takes the scattering amplitude for a dipole of size\   $K$\  on the nucleus. Thus 
\begin{center}
\begin{figure}
\epsfbox[0 0 171 177]{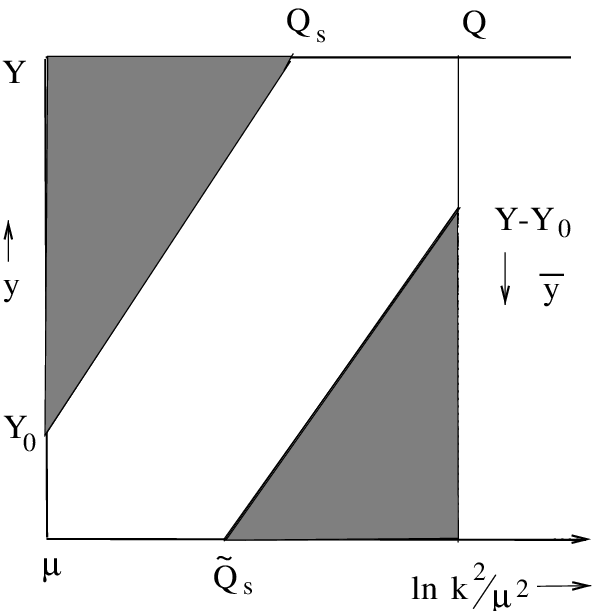}
\caption{}
\end{figure}
\end{center}
\begin{equation}
T_A(\mu,b,Q) \sim \int_0^Y dy\ \int_\mu^Q{dN_g(Q,y,K)\over dy} \cdot t_A(K,b,\mu){dK\over K}
\end{equation} 

\noindent where

\begin{equation}
{dN_g(Q,y,K)\over dy}=\int d^2b^\prime\  {d\tilde{N}_g(Q,b^\prime,y,K)\over dy}.
\end{equation}

\noindent The rapidity of the dipole $K$ in (19) should be greater than $\ell n A^{1/3}$ so that the dipole is coherent over the size of the nucleus, but we suppose that  $\ell n A^{1/3}$ is small enough that it can be neglected in setting the upper limit of the $y-$integral in (19).  Now

\begin{equation}
t_A(K,b,\mu) \sim \alpha^2(1/K^2) \ell n\  
K^2/\mu^2 \cdot \rho 2{\sqrt{R^2-b^2}}
\end{equation}

\noindent so long as $t_A$ is small, where we have taken the gluon distribution of a  dipole $\mu$ in the nucleus to be $\alpha\  \ell n\  K^2/\mu^2$ at scale $K.$  Using (14), (20), and (21) in (19) one finds that the $K$ integration (19) diverges in the infrared.  But, this integration should be cut off at the value, $K_0,$ when $t_A,$ given in (21) reaches one[21-24].  Thus

\begin{equation}
T_A\sim {1\over \alpha^2} ({K_0^2\over \tilde{Q}_s^2})^{1-\lambda_0}\ell n (\tilde{Q}_s^2/K_0^2)
\end{equation}

\noindent where, from (21),

\begin{equation}
{K_0^2\over \mu^2} \equiv {Q_s^2(MV)\over \mu^2} \sim \alpha^2 \ell n (K_0^2/\mu^2) \rho 2{\sqrt{R^2-b^2}}\ /\mu^2.
\end{equation}

\noindent or

\begin{equation}
{K_0^2\over \mu^2} \sim \alpha^2 c A^{1/3} \ell n (\alpha^2 c A^{1/3})
\end{equation}

\noindent with

\begin{displaymath}
2\rho{\sqrt{R^2-b^2}} /\mu^2 = c(b,\mu) A^{1/3}.
\end{displaymath}

\noindent ($K_0^2$ is the \underline{quark} saturation momentum in the Mclerran-Venugopalan model of a nucleus made of dipoles of size $1/\mu.$)  We shall always assume that $\alpha^2c A^{1/3} \gg 1$ in order to have non-trivial nuclear effects.

Now use (24) to eliminate $K_0$ in (22).  This gives

\begin{displaymath}
T_A\sim[\alpha^2cA^{1/3}\ell n(\alpha^2c A^{1/3}){Q_s^2\over Q^2}]^{1-\lambda_0}{1\over \alpha^2}
\cdot\ \ [\ell n {Q^2\over Q_s^2}-\ell n (\alpha^2c A^{1/3}) - \ell n \ell n (\alpha^2c A^{1/3})]
\end{displaymath}

\noindent or

\begin{equation}
T_A \sim ({Q_s^2Q_s^2(MV)\over \mu^2Q^2})^{1-\lambda_0}{1\over \alpha^2} \ell n[{\mu^2Q^2\over Q_s^2Q_s^2(MV)}]
\end{equation}

\noindent The saturation momentum for the nucleus is defined as the value of $Q^2$ at which $T_A$ becomes of order one. This gives

\begin{equation}
\ell n({Q_s^2(A)\over \mu^2}) = \ell n ({Q_s^2\over \mu^2}) + \ell n ({Q_s^2(MV)\over \mu^2}) + {1\over 1-\lambda_0} [\ell n {1\over ^2} + \ell n \ell n 1/\alpha^2] + const
\end{equation}

\noindent where $Q_s^2(MV)$ is the saturation momentum in the McLerran-Venugopalan model.  We note however that $T_A$ does not quite have the usual scaling form (8) since

\begin{equation}
T_A \sim \left({Q_s^2(A)\over Q^2}\right)^{1-\lambda_0}\left(1+ {\ell n(Q^2/Q_s^2(A)\over \ell n[{\mu^2Q_s^2(A)\over Q_s^2 Q_s^2(MV)}]}\right).
\end{equation}

\noindent Eq.(27) is only valid for $Q^2/Q_s^2(A) > 1.$ When $Q^2/Q_s^2(A) < 1$ there is more than one dipole of scale $K_0$ in the parent dipole $Q$ and unitarity effects will further suppress the $T_A$ given in (27)[22-24].  One can express the scaling behavior of $T_A$ more clearly by defining ariable

\begin{equation}
\bar{Q}_s^2(A) = Q_s^2c \alpha^2 A^{1/3} \ell n (c\alpha^2A^{1/3}) = Q_s^2Q_s^2(MV)/\mu^2
\end{equation}  

\noindent in terms of which, from (25),

\begin{equation}
T_A\sim{1\over \alpha^2}\left({\bar{Q}_s^2(A)\over Q^2}\right)^{1\lambda_0} \ell n\left(Q^2/\bar{Q}_s^2(A)\right)
\end{equation}

\noindent where, now,(29) can be used when

\begin{equation}
{Q^2\over \bar{Q}_s^2(A)} > \left({1\over \alpha^2} \ell n {1\over \alpha^2}\right)^{1\over 1-\lambda_0}.
\end{equation}

\noindent While $\bar{Q}_s(A)$  gives the simpler looking scaling behavior it is $Q_s(A)$ which is the actual saturation momentum of the nuclear light-cone wavefunction.

Finally, we evaluate $T_A/T_\mu.$  From (8) and (29) one finds

\begin{equation}
T_A/T_\mu\sim {1\over \alpha^2}\left({\bar{Q}_s^2(A)\over Q_s^2}\right)^{1-\lambda_0}{\ell n(Q^2/\bar{Q}_s^2(A))\over \ell n(Q^2/Q_s^2)}.
\end{equation}

\noindent Using (28) gives 

\begin{equation}
T_A/T_\mu \sim {cA^{1/3}\over (c\alpha^2A^{1/3})^{\lambda_0}} \ell n^{1-\lambda_0}(c\alpha^2A^{1/3})\left(1-{\ell n(c\alpha^2A^{1/3}\over \ell n(Q^2/Q_s^2)}\right).
\end{equation}

\noindent Thus not too far from the saturation boundary there is significant shadowing with the A-dependence of $T_A$ proportional to $(A^{1/3})^{1-\lambda_0}.$  It is this shadowing which, according to Ref.11, causes particle production in heavy ion collisions to scale,roughly, like $N_{part}$.  We note that far from the scaling region, when  ${\ell n Q^2/\mu^2\over \alpha Y} >> 1,$  BFKL dynamics will replace $\lambda_0$ by $0,$ and an $A^{1/3}$ behavior will again emerge for $T_A.$  This is the region where double leading logarithmic behavior is valid.

\section{Running coupling BFKL dynamics}

In this section we revisit our discussion in the last section, but now using running coupling BFKL dynamics.  We shall see that saturation looks quite different when running couplings effects are present.

We begin by considering the scattering of  a dipole $Q$ on a dipole $\mu$ where, as before,  $Q/\mu \gg 1.$  As in the last section[16-17]

\begin{equation}
T_\mu(\mu,b,Q)\simeq T_0\left({Q_s^2\over Q^2}\right)^{1-\lambda_0}\left[\ell n Q^2/Q_s^2 + {1\over 1-\lambda_0}\right].
\end{equation}

\noindent $T_0$ is still a function of $b\mu$ but now $Q_s$ is given by

\begin{equation}
\ell n [Q_s^2/\Lambda^2] = {\sqrt{{4N\chi(\lambda_0)\over \pi b(1-\lambda_0)}\ Y}} + {3\over 4}\left ({a\over c}\right)^{1/3}\xi_1 Y^{1/6}-{1\over 1-\lambda_0}
\end{equation}

\noindent where

\begin{equation}
a={\sqrt{{N_c(1-\lambda_0)[\chi^{\prime\prime}(\lambda_0)]^2\over 4\pi b \chi(\lambda_0)}}}
\end{equation}

\noindent and

\begin{equation}
c={1-\lambda_0\over 2}
\end{equation}

\noindent with $b={11N_c-2N_f\over 12\pi},$ and where $\Lambda$ is the usual QCD parameter.  $\xi_1$ is the first zero of the Airy function, $A_i(\xi).$  Of course, the form given in (33) is valid only when $\ell n Q^2/Q_s^2 \gg 1$ so that the constant terms in (33) and (34) are arbitrary, and unimportant.

The remarkable thing about (34) is that there is no $\mu-$dependence present[16].  As noted earlier the $\mu-$ dependent corrections to (34) can be of size ${\ell n^2(\mu^2/\Lambda^2)\over {\sqrt{Y}}}$ so that in case $\mu^2/\Lambda^2\gg 1$ there is a transition region between a low-$Y$ region where fixed coupling dynamics occurs and a high-$Y$ regime where running coupling dynamics occurs.  This transition value of $Y$ is[16] $Y_{trans}\simeq {\pi(1-\lambda_0)\over 2bN_c\chi(\lambda_0)}\ {1\over \alpha^2(\mu)}.$  Eqs. (33) and (34) apply well above this transition regime.  The lack of a $\mu-$dependence in $Q_s$ at large  $Y$ indicates the insensitivity of $Q_s$ to the nature of the target probed by the dipole $Q.$  Thus, (33)-(36) apply equally to a proton as to a dipole.  And, of course,these equations must apply also to nuclei indicating that $Q_s$, at large $Y,$ has no $A-$dependence.  This is in striking contrast to $A^{1/3}$-dependence of $Q_s^2$ found both in the McLerran-Venugopalan model and in fixed coupling BFKL evolution as given by (6) and (26), respectively.  It is the purpose of the present section to try and explain why there is no $A-$ dependence in $Q_s^2$ in running coupling evolution.

It is useful to consider an expression for $\ell n Q_s^2/\Lambda^2$ which interpolates between fixed coupling evolution and running coupling evolution.  To that end consider

\begin{equation}
\rho_s(Y,\mu) = {\sqrt{{4N_c \chi(\lambda_0)\over \pi b(1-\lambda_0)}Y + \ell n^2(\mu^2/\Lambda^2)}} .
\end{equation}

\noindent When $Y/\ell n^2(\mu^2/\Lambda^2)\ll 1$

\begin{equation}
\rho_s\simeq \ell n(\mu^2/\Lambda^2) + {2\alpha(\mu)N_c \chi(\lambda_0)\over \pi (1-\lambda_0)}\ Y
\end{equation}

\noindent while when $Y/\ell n^2(\mu^2/\Lambda^2) \gg 1$

\begin{equation}
\rho_s \simeq {\sqrt{{4N_c \chi(\lambda_0)\over \pi b(1-\lambda_0)}Y}} + {\ell n^2(\mu^2/\Lambda^2)\over 2{\sqrt{{4N_c \chi(\lambda_0)\over \pi b(1-\lambda_0)}Y}}}
\end{equation}

\noindent so that $\rho_s$ matches onto the \underline{dominant parts} of $\ell n(Q_s^2/\Lambda^2)$ in both the fixed coupling regime, (38), and the running coupling regime, (39).  Now view the evolution running backwards, starting at a large scale $Q$ and ending up at a smaller scale. As in the previous section call $\tilde{Q}_s$ the boundary of the saturation region in the backward evolution.  Then in the approximation (39), and with $\tilde{\rho}_s = \ell n \tilde{Q}_s^2/\Lambda^2,$ one has

\begin{equation}
\tilde{\rho}_s^2=2 {\sqrt{{4N_c \chi(\lambda_0)\over \pi b(1-\lambda_0)}Y}}\left(\rho - {\sqrt{{4N_c \chi(\lambda_0)\over \pi b(1-\lambda_0)}Y}}\right)
\end{equation}

\noindent where $\rho = \ell n Q^2/\Lambda^2.\ \  \tilde{\rho}_s$ is, of course, a function of $Q^2$ and $Y.$  Eq.(40) gives a property of the light-cone wavefunction of a dipole $Q$ and may be applied to the scattering on any target, at least so long as $\tilde{\rho}_s$ remains in the perturbative regime.

Begin by applying (40) to a large nucleus.  Then, from (23), the scattering amplitude is at the edge of the unitarity limit (the edge of the saturation region) when

\begin{equation}
\tilde{\rho}_s(A) = \ell n\left({Q_s^2(MV)\over \Lambda^2}\right)
\end{equation}

\noindent where now $\Lambda^2$ replaces $\mu^2$ in the running coupling case and for a realistic nucleus.  From (40)

\begin{equation}
\ell n^2\left({Q_s^2(MV)\over \Lambda^2}\right)=2{\sqrt{{4N_c \chi(\lambda_0)\over \pi b(1-\lambda_0)}Y}}\left(\ell n({Q_s^2(A)\over \Lambda^2})-{\sqrt{{4N_c \chi(\lambda_0)\over \pi b(1-\lambda_0)}Y}}\right)
\end{equation}

\noindent determines $Q_s^2(A).$  Now consider the scattering on a proton.  Strictly speaking (40) does not apply since the proton is in the nonperturbtive regime. But, it is clear that the unitarity limit, for a fixed impact parameter, will be reached when $\tilde{Q}_s\simeq \Lambda,$ that is when dipoles of size $1/\Lambda$ appear with high probbility in the wavefunction of the dipole $Q.$  Thus

\begin{equation}
O=2{\sqrt{{4N_c \chi(\lambda_0)\over \pi b(1-\lambda_0)}Y}}\left(\ell n({Q_s^2(P)\over \Lambda^2})-{\sqrt{{4N_c \chi(\lambda_0)\over \pi b(1-\lambda_0)}Y}}\right)
\end{equation}

\noindent determines $Q_s^2 $\ for the proton.  Comparing (42) and (43)

\begin{equation}
\ell n{Q_s^2(A)\over Q_s^2(P)}\simeq {Q_s^2(A)-Q_s^2(P)\over Q_s^2(P)}\simeq{\ell n^2({Q_s^2(MV)\over \Lambda^2})\over 2{\sqrt{{4N_c\chi(\lambda_0)\over \pi b(1-\lambda_0)}Y}}}\simeq{\ell n^2({Q_s^2(MV)\over \Lambda^2})\over 2\ell n(Q_s^2/\Lambda^2)},
\end{equation}

\noindent where again we emphasize that $Q_s(MV)$ is the \underline{quark} saturation momentum as determined in the McLerran-Venugopalan model.  Thus at large  $Y$ we see that there is no distinction between $Q_s(A)$ and $Q_s(P)$ although at current energies and for large nuclei the fixed coupling regime may be appropriate.

\end{document}